\documentclass[twoside]{scrartcl}
\usepackage[T1]{fontenc}
\usepackage[latin9]{inputenc}
\usepackage{units}
\usepackage{array}
\usepackage{graphicx}
\usepackage{epstopdf}
\usepackage{float}
\usepackage{amsmath}
\usepackage{amssymb}
\usepackage{esint}
\usepackage{multicol}
\usepackage{changepage}

\usepackage{xcolor}
\usepackage{float}
\usepackage{placeins}

\usepackage[justification=justified]{caption}
\usepackage{prettyref}
\newrefformat{sfig}{Supp. Fig. \ref{#1}}
\newrefformat{tab}{Supp. Table \ref{#1}}
\DeclareCaptionLabelFormat{supptable}{#1 S#2} 
\DeclareCaptionLabelFormat{suppfig}{#1 S#2} 

\makeatletter

\pdfpageheight\paperheight
\pdfpagewidth\paperwidth

\newcommand{\tcontact}{\ensuremath{t_{\mathrm{contact}}}}
\newcommand{\tfusion}{\ensuremath{t_{\mathrm{fusion}}}}
\newcommand{\xinitial}{\ensuremath{x_{\mathrm{initial}}}}
\newcommand{\xcontact}{\ensuremath{x_{\mathrm{contact}}}}

\newcommand{\V}{\ensuremath{v_{\mathrm{front}}}}
\newcommand{\Sig}{\ensuremath{\sigma_{\mathrm{front}}}}
\newcommand{\dfront}{\ensuremath{d_{\mathrm{interface}}}}
\newcommand{\VF}{\ensuremath{v_{\mathrm{interface}}}}

\makeatother

\begin{document}

\begin{flushleft}
%\newline
{\Large
\textbf{Collective stresses  
drive competition between monolayers of normal and Ras-transformed cells
}
}
\newline
\\
Sarah Moitrier\textit{$^{abc\ddag}$}, Carles Blanch-Mercader\textit{$^{d\ddag}$}, Simon Garcia\textit{$^{abc}$}, Kristina Sliogeryte\textit{$^{abc}$}, Tobias Martin\textit{$^{abc}$}, Jacques Camonis\textit{$^{ef}$}, Philippe Marcq\textit{$^{ab*}$}, Pascal Silberzan\textit{$^{abc}$} and Isabelle Bonnet\textit{$^{abc*}$}
\\
\bigskip
% \\
$^{a}$~Laboratoire Physico Chimie Curie, Institut Curie, PSL Research University, CNRS UMR168, 75005 Paris, France\\
$^{b}$~Sorbonne Universit\'e, 75005, Paris, France\\
$^{c}$~\'Equipe labellis\'ee Ligue Contre le Cancer\\
$^{d}$~Universit\'e de Gen\`eve, Geneva, Swizterland\\
$^{e}$~Institut Curie, PSL Research University, 75005 Paris, France\\
$^{f}$~ART group, Inserm U830, 75005 Paris, France\\
$^{\ddag}$ These authors contributed equally to this work\\ 
$^*$ Authors for correspondence: \texttt{isabelle.bonnet@curie.fr}, \texttt{philippe.marcq@curie.fr}
\end{flushleft}

%\date{}

%\newpage
\setcounter{section}{0}
\setcounter{figure}{0}
\setcounter{table}{0}
\setcounter{equation}{0}

\section*{Abstract}

We study the competition for space between two cell lines that differ 
only in the expression of the Ras oncogene. The two cell populations 
are initially separated and set to migrate antagonistically towards an 
in-between stripe of free substrate. After contact, their interface 
moves towards the population of normal cells. We interpret the velocity and traction 
force data taken before and after contact thanks to a hydrodynamic  
description of collectively migrating cohesive cell sheets. The kinematics of cells, before and after contact, allows us to estimate the relative material parameters for both cell lines.
As predicted by the model, the transformed cell population with 
larger collective stresses pushes the wild type cell population.

\section{Introduction}
Living organisms are composed of several tissues where cells continuously interact and compete for resources and 
space to ensure tissue cohesion and functionality \cite{Beco2012,Levayer2013,Amoyel2014,Merino2016}. Competitive interactions lead to the elimination of non-optimal cells and are crucial to maintain tissue integrity, homeostasis and function. Tissue organization is extremely stable but can be compromised in pathological situations, for instance in the case of tumor proliferation, where 
competitive cell interactions may also play a role \cite{Eichenlaub2016}. Strikingly, it has indeed been proposed that 
precancerous cells could act as supercompetitors killing surrounding cells to make room for themselves \cite{Moreno2004}. 
Conversely, it has been observed that isolated cells either carrying tumor-promoting mutations \cite{Hogan2009,Gullekson2017,Wagstaff2016} or deprived of tumor-suppressor genes \cite{Norman2012}, are eliminated from the wild 
type tissue. Importantly, the properties of entire groups of cells go beyond the sum of those of individual cells. A comprehensive understanding of these effects requires to integrate cell-cell interactions over tissue scales.

\noindent Recently, confrontation assays between antagonistically migrating cell 
sheets have been used \cite{Nnetu2012,Porazinski2016,Taylor2017,Rodriguez-Franco2017}, in particular  to study the interactions 
between normal and GFP-Ras$^{V12}$ Madin-Darby canine kidney (MDCK) cells \cite{Porazinski2016}.
When Ras$^{V12}$ and normal cells meet, the Ras$^{V12}$ cells collapse and 
are displaced backwards, while normal cells continue to migrate forward. This displacement 
of the interface does not rely on the classical principle of contact inhibition of locomotion. From a biological point of view, it 
has been attributed to an ephrin-dependent mechanism: 
normal cells detect transformed Ras$^{V12}$ cells through interactions 
between ephrin-A and its receptor EphA2. 
Using similar confrontation assays between two cell types expressing the 
EphB2 receptor and its ligand ephrinB1, it has been further shown
that the repulsive interactions between two cell types drives cell segregation 
and border sharpening more efficiently than a low level of heterotypic adhesion
\cite{Taylor2017}. The mechanical interactions between two populations may
lead to  oscillatory traction force patterns, which pull cell-substrate 
adhesions away from the border, and may trigger deformation waves,
generated at the interface between the two cell types and propagating 
across the monolayers \cite{Rodriguez-Franco2017}. The biomechanical 
determinants of dominance of a given cell population over another one remain
unclear, as different theoretical descriptions of cell competition 
rely on differences in proliferation rates \cite{Ranft2014},
in cell motilities \cite{Lorenzi2016,Hallou2017}, or predator-prey interactions 
\cite{Nishikawa2016}.

\noindent In this work, we investigate the mechanisms of 
competitive cell interactions between normal and precancerous Human Embryonic Kidney (HEK) cell assemblies. 
In particular, we assess the invasive capacity of oncogene-bearing cells by 
adapting the classical wound healing assay \cite{Poujade2007} to an antagonistic migration 
assay (AMA) of two cell populations 
\cite{Nnetu2012,Porazinski2016,Taylor2017,Rodriguez-Franco2017}. This approach 
holds the advantage of creating an interface between 
two cell populations in a reproducible way. Each cell type is seeded 
into one of the compartments 
of a cell culture insert so as to be initially separated by a gap.
When the culture insert is removed, cells migrate to close the gap, and 
facing cells eventually meet. Later, it is observed that the \emph{transformed} cell sheet 
penetrates the spatial domain occupied by the wild type cell sheet and 
displaces it backwards. We adapt a biophysical model previously introduced
\cite{Blanch-Mercader2017} to describe the early dynamics of
expansion of a single cell sheet into cell-free space and extract mechanical parameter values. Comparing theoretical 
predictions with experimental data,
we show that differences in the amplitude of collective
stresses developed at the free edges of the two independent migrating monolayers explain the displacement of the wild type cell population 
by the transformed cell sheet.

\section{Materials and methods}
\subsection{Cell culture}
Human Embryonic Kidney cell lines have been immortalized by ectopic 
expression of large-T and hTERT genes: the HEK-HT cells \cite{Hahn1999}. From now on, we refer to these cells as "HEK cells". 
In this work, we use the two following cell lines:
\begin{itemize}
\item HEK-GFP: a variant transduced to globally express the green fluorescent
protein GFP, referred to below as the ``wild type'' or 
``normal'' cell line (HEK wt);
\item HEK-Ras-mCherry: a cell line carrying the H-RAS$^{G12V}$ 
mutation, and transduced to globally express the fluorescent 
protein mCherry, referred to below as the ``Ras'' or 
``transformed'' cell line (HEK Ras).
\end{itemize}

\noindent Cells were cultured in Dulbecco's Modified Eagle's Medium 
(DMEM GlutaMAX, Gibco) supplemented with penicillin-streptomycin (Gibco) 
and fetal bovine serum (FBS, Gibco) - respectively 1\% and 10\% vol/vol - 
at 37$^{\circ}$C, 5\% CO$_2$ and 95\% relative humidity. The medium was also 
supplemented with selection antibiotics according to cell line's 
specific resistance, namely with hygromycin B ($100 \,\mu \mathrm{g/mL}$,
Gibco) and geneticin ($400 \,\mu \mathrm{g/mL}$, Gibco) for both cell lines, 
and with additional puromycin ($0.5 \,\mu \mathrm{g/mL}$, Gibco) for the 
Ras cell line.

\subsection{Population doubling time}
For the estimation of the population doubling time $\tau_d$, cells from each cell line were seeded in 8 wells of a plastic bottom 24-well plate. Twice a day, for 4 consecutive days, the cells from one well were resuspended using Trypsin-EDTA (Gibco) and counted in a given volume using a KOVA Glasstic Slide 10 with Grids (KOVA). Assuming that the number $n$ of cells as a function of time after seeding follows $n(t) = n_0 2^{t/\tau_d}$, where $n_0$ denotes the initial cell number, we deduce an estimation from the slope of the graph $n = f(t)$ with semi-logarithmic axes since $\log(n) = \log(n_{0})+t \log(2)/\tau_d$. We found $\tau_d^{\mathrm{wt}} = 16 \pm 3$h and $\tau_d^{\mathrm{Ras}} = 16 \pm 1$h, see Fig.~S1. % Fig.~S1.

\subsection{Immunostaining}
Cells were fixed using 4\% paraformaldehyde (PFA, Electron Microscopy Science, ref. 15710) for 20~min. Samples were then washed three times in phosphate buffer saline (PBS). For permeabilization, cells were treated with 0.5\% Triton X-100 for 10~min, followed with three rinsing steps in PBS. Non-specific binding was blocked by incubating in 3\% bovine serum albumine (BSA, Sigma) in PBS for 30~min. Samples were then incubated with primary antibody N-cadherin rabbit (7939, Santa Cruz) diluted 1:200 and E-cadherin mouse (610181, BD biosciences) diluted 1:100 in PBS with 0.5\% BSA for 60~min. After incubation, samples were washed three times in PBS and incubated in secondary antibody Alexa Fluor 488 chicken anti-rabbit and Alexa Fluor 546 goat anti-mouse (respectively A21441 and A11003, both from Invitrogen) diluted in 1:1000. DNA binding dye (DAPI, ThermoFisher) was added at $1~\mu$g.mL$^{-1}$ in PBS with 0.5\% BSA for 60~min. The samples were washed again in PBS and mounted with Prolong Gold reagent (Life technologies). Images were acquired with a Zeiss LSM NLO 880 confocal microscope using ZEN software.  The final images are presented as the sum of Z-stacks. We used MDCK cells as a control for antibody validation.

\subsection{Antagonistic migration assay}
\noindent We used commercially available silicone-based Culture-Inserts 
2 Well (Ibidi), whose outer dimensions are $9 \times 9 \, \mathrm{mm}^2$. 
Each well covers a surface of $22 \, \mathrm{mm}^2$. The insert was placed in 6-well glass bottom plates
(IBL, Austria) and the cells were seeded at roughly 0.5 million cells/mL. The normal cell type was always seeded in the left compartment of the culture insert, while the transformed cells were seeded in the right compartment. Cells were left to incubate overnight until fully attached - then, the culture insert was removed, leaving a free space between the two monolayers, which could then migrate towards each other to close this gap.

\begin{figure}[ht!]
	\centering
	\includegraphics[height=10cm]{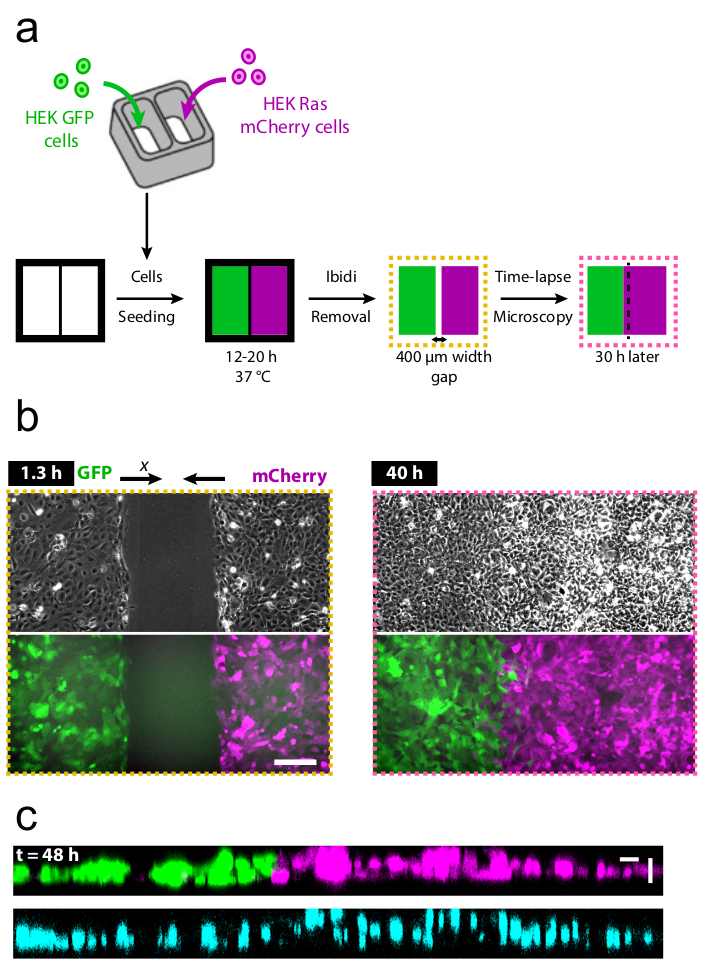}
	\caption{\textbf{Principle of the antagonistic migration assay.} 
		a: Schematics of the experimental procedure. The time reference $t=0$ corresponds to insert removal.
		b- Left: initially, the two cell populations are separated by a cell-free gap of $\Delta x \sim 400 \,\mu \mathrm{m}$ . 
		Phase contrast (top) and fluorescence images (bottom) of the cell 
		monolayers at $t = 1.3$ h. The cell populations 
		migrate along the $x$ axis in opposite directions. 
		b- Right: gapless monolayer at $t = 40$ h.
		Scale bar: $200 \,\mu\mathrm{m}$.
		c: Side view of the AMA at $t = 48$ h (not the same experiment as in b). Top: HEK-GFP  cells (green) and HEK-Ras-mCherry cells (magenta). Bottom: Hoechst labelling (nuclei). Note that after fusion,
                the tissue remains monolayered. Scale bars: $25 \,\mu\mathrm{m}$}
	\label{fig01}
\end{figure}

\noindent  The plane occupied by the cell sheets is described by cartesian
coordinates, where $x$ denotes the direction of migration, see Fig.~\ref{fig01}.
Initially, the two monolayers are set apart by a cell-free gap of width 
$\Delta x = |\xinitial^{\mathrm{wt}}-\xinitial^{\mathrm{Ras}}| \sim 400 \,\mu \mathrm{m}$.  
The removal of the barrier sets the reference time $t=0$. 

\noindent Time-lapse experiments were carried out using an 10x objective 
(HCX PL Fluotar, 0.3 Ph1, Leica) mounted on an DM-IRB inverted microscope 
(Leica) equipped with temperature, humidity, and CO$_2$ regulation 
(Life Imaging Services). The motorized stage (H117 motorized stage, 
Prior Scientific), and the image acquisition with a CCD camera 
(CoolSnap EZ (Photometrics) or Retiga 6000 (Qimaging)) 
were controlled using Metamorph software (Molecular Devices). 
The typical delay between successive images was $15$ min.  We followed the AMA during 3 days by acquiring 
images in three channels: phase contrast (all cells), GFP (HEK wt cells) 
and mCherry (HEK Ras cells). In this work, the analysis has been performed during the first $60$ h after barrier removal, that is until the tissue becomes multilayered.

\noindent Custom-made ImageJ \cite{Rasband2012} macros were used to automatically process large numbers of
 images for stitching, merging channels and assembling movies.
 We have used green and magenta as false colours for GFP and mCherry signals.

\subsection{Velocimetry}
The velocity fields in the cell monolayers were analyzed by particle image velocimetry 
(PIV) using the MatPIV toolbox for Matlab (Mathworks), as previously described 
\cite{Petitjean2010,Deforet2012}. The window size was set to 16 pixels ($\sim 40 \,\mu \mathrm{m}$ typically), 
with an overlap of 0.25. Sliding average over 1~h was performed.

\noindent Averaging the velocity fields along the $y$ direction, we fitted velocity profiles $v(x)$ 
with a single exponential function $\sim  V\, \exp((x-L(t))/\lambda)$, 
where $L(t)$ is the position of the front at time $t$ and it is determined by the position of the extrema of the measured velocity profiles.
To improve accuracy of the measurement, the parameters $V$ and $\lambda$ 
were estimated from the first two moments of the velocity profiles,
as it led to substantially smaller error bars than other fitting procedures. For instance, our estimates of parameters of the normal monolayer
velocity profile read $\lambda^{\mathrm{wt}} = 
\int_{-\infty}^{L^{\mathrm{wt}}(t)}(L^{\mathrm{wt}}(t)-x) v(x) \, \mathrm{d}x/
\int_{-\infty}^{L^{\mathrm{wt}}(t)}v(x) \, \mathrm{d}x$,
$V ^{\mathrm{wt}} = \int_{-\infty}^{L^{\mathrm{wt}}(t)}v(x) \, \mathrm{d}x
/ \lambda^{\mathrm{wt}}$. Similar expressions can be derived to estimate 
$\lambda^{\mathrm{Ras}}$ and $V^{\mathrm{Ras}}$.

\subsection{Traction Force Microscopy}
We adapted the protocol from Tse and Engler \cite{Tse2010}. First, we prepared "activated" coverslips : coverslips were cleaned in a plasma cleaner for 10 minutes, incubated in a solution of 3-aminopropyltrimethoxysilane (2\% vol/vol in isopropanol, Sigma) for 10 minutes, and rinsed with distilled water. They were then incubated in glutaraldehyde (0.5\% vol/vol in water, Sigma) for 30 minutes, and air dried. Independently, microscope glass slides were incubated in a solution of Fibronectin Bovine Protein (Gibco) in PBS at $25 \, \mu \mathrm{g/mL}$ for 30 minutes, then left to air dry. We mixed a solution of 40\% acrylamide (Bio-Rad) with a solution of 2\% bis-acrylamide (Bio-Rad) in water, and added 1\% (vol/vol) of fluorescent beads (FluoSpheres $0.2 \,\mu \mathrm{m}$ dark red fluorescent 660/680, Life Technologies) in order to make a gel of $\sim 20  \mathrm{kPa}$.

\noindent To start the polymerization of the gel, ammonium persulfate (1\% ~vol/vol, Bio-Rad) and TEMED (1\textperthousand ~vol/vol Bio-Rad) were added to the solution containing the beads and thoroughly mixed. Then $30 \,\mu\mathrm{L}$ was applied on the fibronectin-coated slides, and activated coverslips were placed on top. This step, inspired by the deep-UV
patterning technique \cite{Azioune2010}, enabled us to directly coat the surface of the gel with fibronectin. 

\noindent When the polymerization was complete, the sandwiched gel was immersed in PBS, and the coverslip bearing the gel was carefully detached. It was then incubated in culture medium for 45 minutes, at $37^{\circ}$C, before the cells were seeded on its surface, and left to adhere overnight. We finally used a POCmini-2 cell cultivation system (Pecon GmbH) for image acquisition under the microscope. The images were acquired as usual, with the added far red channel to image the beads. Reference images of the beads in the gel at rest were taken after trypsinization. Traction forces were computed using the Fiji plugins developed by Tseng \emph{et al.} \cite{Martiel2015}. 

\noindent Note that TFM experiments are done on soft acrylamide gels, which are fibronectin-coated, while we generally carried out experiments on plain glass.  However, experiments conducted on fibronectin-coated glass showed that fibronectin does not change the final outcome of the AMA, although it may affect its dynamics. The traction force measurements are acquired $t=1$~h after barrier removal 
at uniform time intervals of 15 minutes. To improve the accuracy of our data, 
time averages are performed over time windows of $2.5$~hours. 
We observe a relaxation of the spatial autocorrelation 
function of both components $T_x$ and $T_y$ of the traction force field, 
and estimate the traction force correlation lengths 
by the position at which a linear extrapolation near the maximal value ($x=0$)
of the autocorrelation function crosses the $x$-axis.
For isolated cells, we measure both traction forces
(see Fig.~S2) and strain energy density. The latter
is the strain energy divided by the cell area \cite{Butler2002}. 

\subsection{Statistical analysis}

Statistical significance was quantified by p-values calculated by a t-test (Fig.~\ref{fig06}) or a Mann-Whitney U test (other figures).
Different levels of significance are shown on the graphs: * p $\leq$ 0.1; ** p $\leq$ 0.05;  *** p $\leq$ 0.01. p-values larger than 0.1 were considered not significant ('n.s').

\subsection{Model}
We briefly summarize here the model
of an active viscous material proposed in Blanch-Mercader \textit{et al.} \cite{Blanch-Mercader2017} 
to describe the expansion of a planar cell sheet spreading in a direction
defining the $x$ axis, in the limit where the extension of the system 
along the $y$ axis is much larger than along $x$. In this 
case, approximate translation invariance along $y$ allows to treat the system 
in 1D along the $x$ axis, by averaging all relevant fields over $y$. 
We denote $v(x,t)$, $p(x,t)$ and $\sigma(x,t)$ 
the $x$ components of the velocity, polarity and stress fields. Within a continuum mechanics approach, 
the equations governing monolayer expansion into free space read:
\begin{eqnarray}
  \label{eq:def:model:1:const}
  \sigma &=& \eta \, \partial_x v \,,\\
  \label{eq:def:model:1:balance}
  \partial_x \sigma &=& \xi v - T_0 p \,,\\
  \label{eq:def:model:1:polarity}
  0 &=& p - L^2_c \, \partial^2_x p \,.
\end{eqnarray}
They respectively represent: \eqref{eq:def:model:1:const} 
the constitutive equation for a viscous 
compressible fluid with viscosity $\eta$; \eqref{eq:def:model:1:balance}
the force balance equation at low Reynolds number in the presence of both 
passive (friction coefficient  $\xi$) and active (magnitude $T_0$) 
traction forces; and \eqref{eq:def:model:1:polarity} the polarity equation 
in the quasi-static limit. The length $L_c$ is the length scale over which 
the monolayer front is polarized and generates active traction forces.

\noindent In the case of a single monolayer located in the domain
$x \le L(t)$ at time $t$, and expanding towards $x>0$, 
the boundary conditions read:
\begin{eqnarray}
  \label{eq:def:BC:1}
  \sigma(x = L(t), t) &=& 0 \,,\\
  p(x = L(t), t) &=& +1 \,,
\end{eqnarray}
leading to the polarity profile ($x \le L(t)$):
\begin{equation}
  \label{eq:sol:p}
  p(x, t) = \exp \left( \frac{x - L(t)}{L_c}\right)
\end{equation}
and to the velocity and stress profiles:
\begin{eqnarray}
  \label{eq:sol:v:1}
  v(x, t) &=& \frac{ \V}{L_c - L_{\eta}} 
\left(
L_c \, \exp \left( \frac{x - L(t)}{L_c}\right)
- L_{\eta} \, \exp \left( \frac{x - L(t)}{L_{\eta}}\right)
\right)\,, \\
\sigma(x, t) &=&\frac{\eta \V}{L_c - L_{\eta}} 
\left(\exp \left( \frac{x - L(t)}{L_c}\right)
- \exp \left( \frac{x - L(t)}{L_{\eta}}\right)
\right)\,,
\end{eqnarray}
where  
\begin{equation}
 \label{eq:def:Ln}
L_{\eta} = \sqrt{\eta/\xi}
\end{equation}
is the hydrodynamic length and the 
velocity $\V$ of the moving front reads:
\begin{equation}
  \label{eq:def:vfront:1}
   \V =v(x = L(t), t) = \frac{T_0 L_c}{\xi(L_c + L_{\eta})} \,.
\end{equation}

\begin{figure}[ht!]
	\centering
	\includegraphics[height=8cm]{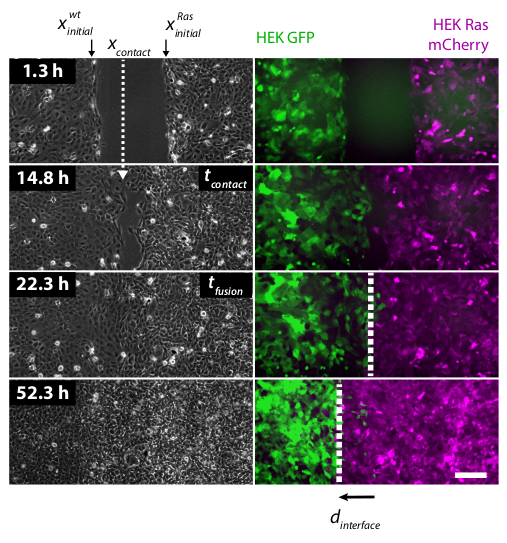}
	\caption{\textbf{Kinematics of the antagonistic migration assay
			between HEK-GFP (green) and HEK-Ras-mCherry (magenta) cells.} 
		 Four time-points of a representative AMA. Phase contrast (left) and fluorescence images (right).The time reference $t=0$ corresponds to the barrier's removal. From top to bottom: first time-point of the acquisition, \tcontact, \tfusion ~and $t= 30$~h after the fusion. 		
		Note the backward migration of the HEK-GFP population after contact and fusion. We indicate $\xinitial^{\mathrm{wt}}$, 
		$\xinitial^{\mathrm{Ras}}$, $\xcontact$ (white dashed line), as well as the displacement \dfront ~of the interface within
		$30$ h after $\tfusion$.  Scale bar: $200 \,\mu\mathrm{m}$. }
			\label{fig02}
		\end{figure}

\noindent This model can be generalized to describe the mechanical behavior
of the gapless monolayer after fusion of the expanding normal and transformed
cell sheets. With the geometry of Fig~\ref{fig01} in mind, we denote quantities 
pertaining to the transformed (respectively normal) cells with the 
index $r$ (respectively $l$), occupying the domain defined by
$x \ge L(t)$ (respectively $x \le L(t)$) at time $t$.

\noindent Eqs.~(\ref{eq:def:model:1:const}-\ref{eq:def:model:1:polarity}) 
apply for each cell sheet, distinguished by a set of distinct parameters:
\begin{eqnarray}
  \label{eq:def:model:2:const}
  \sigma^{l,r} &=& \eta^{l,r} \, \partial_x v^{l,r}. \,\\
  \label{eq:def:model:2:balance}
  \partial_x \sigma^{l,r} &=& \xi^{l,r} v^{l,r} - T_0^{l,r} p^{l,r} \,,\\
  \label{eq:def:model:2:polarity}
  0 &=&  p^{l,r} - (L^{l,r}_c)^2 \, \partial^2_x p^{l,r} \,.
\end{eqnarray}
The boundary conditions at the interface read:
\begin{eqnarray}
  \label{eq:def:BC:2:sig}
  \sigma^l(x = L(t), t) &=& \sigma^r(x = L(t), t) \,,\\
  v^l(x = L(t), t) &=& v^r(x = L(t), t) \,,\\
  \label{eq:def:BC:2:pl}
  p^l(x = L(t), t) &=& 1 \,,\\  
  \label{eq:def:BC:2:pr}
  p^r(x = L(t), t) &=& -1 \,.
\end{eqnarray}
An important assumption is that we ignore a possible repolarization of the cell sheets after a change of 
the direction of migration, Eqs.~(\ref{eq:def:BC:2:pl}-\ref{eq:def:BC:2:pr}). Integration of the evolution equations 
(\ref{eq:def:model:2:const}-\ref{eq:def:model:2:polarity})
with boundary conditions (\ref{eq:def:BC:2:sig}-\ref{eq:def:BC:2:pr}) 
leads to the following expression of the interface velocity $\VF$:
\begin{equation}
  \label{eq:def:vfront:2}
  \VF = v(x = L(t), t) = \frac{L_{\eta}^r \, \eta^l \, \V^l - L_{\eta}^l \, \eta^{r} \, \V^r}{L_{\eta}^l \, \eta^r + L_{\eta}^r \, \eta^{l}} \,,
\end{equation}
with left and right front velocities $\V^{l,r}$ obtained as above
using \eqref{eq:def:vfront:1} with $l$ and $r$ material parameters. 
Remarkably, the interface velocity can be rewritten as 
\begin{equation}
	\label{eq:def:vfront:3}
	\VF = \frac{\Sig^l - \Sig^r}{\eta^r/L_{\eta}^r + \eta^{l}/L_{\eta}^l} \,,
	\end{equation}
upon defining   
\begin{equation}
\label{eq:def:sfront:1}
 \Sig  = \frac{ \eta \, \V }{L_{\eta}}=   \frac{T_0 L_c L_\eta}{L_c+L_{\eta}} \,,
\end{equation}
where the front stress $\Sig$ can be interpreted as the maximum stress 
value within a cell sheet whose boundary is clamped at a fixed position 
(see Appendix A).
The direction of motion of the interface between two competing tissues,
given by the sign of $\Sig^l - \Sig^r$, is determined by the collective stresses that build up at the fronts.

\begin{figure}[ht!]
	\centering
	\includegraphics[height=5.2cm]{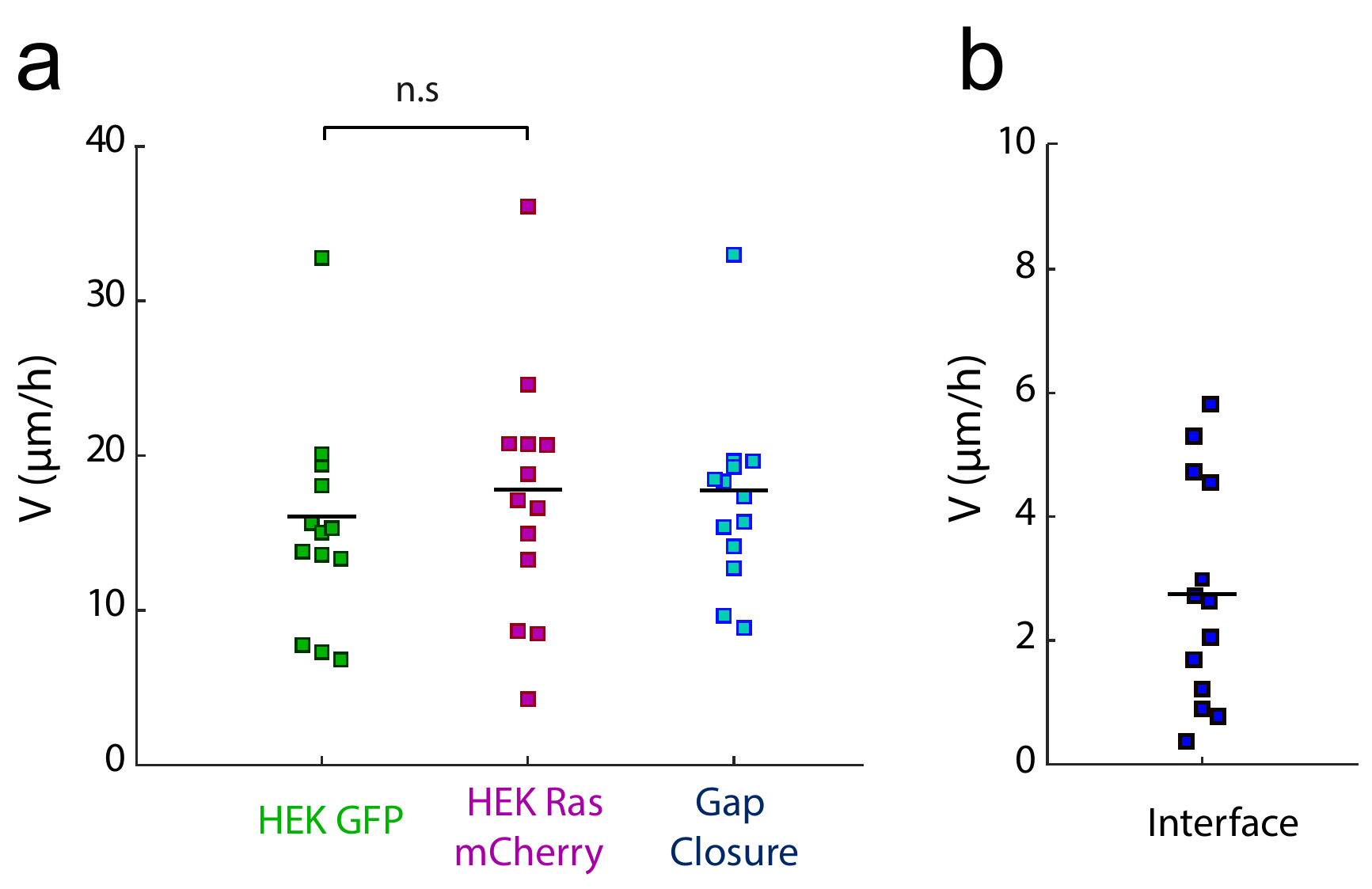}
	\caption{\textbf{Front velocities.} a : velocities of the fronts of 
		expanding cell sheets (HEK-GFP in green and HEK-Ras-mCherry in magenta) and of the gap closure. b: velocity of the interface between the two populations after the meeting. Horizontal black lines correspond to mean values.}
	\label{fig03}
\end{figure}

\section{Results}

\subsection{Characterization of cell lines}
The two HEK cell lines form monolayers in culture (see phase contrast images of Figs~\ref{fig01}-\ref{fig02} + ESI movie). The Ras mutation does not affect 
the population doubling time of the cells since we found a doubling time of 
$16 \pm 3$ h for normal and $16 \pm 1$ h for transformed cells (Fig.~S1).

\noindent  We note that a confluent monolayer of 
transformed cells contains about twice as many cells as a confluent monolayer 
of normal cells for the same area. Indeed, isolated HEK normal cells are approximately twice as large as HEK Ras cells: we 
measured a mean area of $3100 \pm 1100 \,\mu\mathrm{m}^2$ 
and $1600 \pm 400 \,\mu\mathrm{m}^2$ for normal and transformed
cells, respectively (Standard Deviation SD, $n=14$).

\noindent 
In order to mechanically characterize the two cell lines, we first
  estimated the traction forces developed by isolated HEK cells on their
  substrate using traction force microscopy.
We found that the mean traction force amplitude was larger for HEK wt
  cells compared to HEK Ras cells: $110 \pm 32 \,\mathrm{Pa}$ and
  $65 \pm  29 \,\mathrm{Pa}$, respectively  (SD, $n=15$ and $n=13$, Fig.~S2). 
  Since the two cell types differ in size, we also computed the strain energy
  density, and found that the strain energy density was
about 3 times higher for wt cells compared to Ras cells:
$3.5 \pm  2.1 \times 10^{-5} \,\mathrm{J. \,m}^{-2}$ and
$1.1 \pm  0.7 \times 10^{-5} \,\mathrm{J. \,m}^{-2}$
(SD, $n=15$ and $n=13$, not shown). 
Such a decrease of traction forces upon the expression of H-Ras has been reported for isolated NIH3T3 fibroblasts \cite{Munevar2001}. 

\noindent
  Next, we analyzed the statistical properties of collective cell traction forces far from the margin, focusing on two windows of $0.6~\text{mm} \, \times 3~\text{mm}$ on the leftmost region of HEK-GFP monolayers and on the rightmost region
  of HEK-Ras-mCherry monolayers (see Fig.~\ref{fig01}). The leading
  edges were at least $700$ $\mu$m away from the analyzed force data in these windows. Fig.~S3,a shows that the distribution of force orientation
  was approximately uniform for both cell types, suggesting that both
  monolayer subsets were mechanically disconnected from the corresponding
  leading edges, and that possible traction force correlations occur over a length scale smaller than $700 \,\mu$m. Fig.~S3,b shows that normal cells exerted forces
of amplitudes $51\pm 8$~Pa (SD, n=12), comparable to those exerted by transformed cells	$48\pm 3$ (SD, n=12).
Importantly, collective cell traction force behaviour
  could not be extrapolated from single cell traction forces.

\subsection{Before contact, both monolayers migrate freely}
\noindent Upon removal of the insert, the monolayers migrate toward each other, 
while spreading on the free surface. The phase contrast 
images allow us to extract the position $\xcontact$ 
of the first contact between the two opposite populations as well as the 
corresponding time $\tcontact$ (Fig~\ref{fig02} - 2$^{\mathrm{nd}}$ panel). 
We define a second characteristic time: 
$\tfusion$ which is the time when the gap closes completely 
(Fig.~\ref{fig02}- 3$^{\mathrm{rd}}$ panel). We observe that 
the two populations meet at $\tcontact = 16.5 \pm 5.6$ h (SD, $n = 13$) 
after barrier removal, and that the gap closes completely within 
$\tfusion = 27.9 \pm 6.3$ h (SD, $n = 13$). We define the average front 
velocity of each monolayer as: 
$\V^{\mathrm{wt},\mathrm{Ras}}=\frac{|\xcontact-\xinitial^{\mathrm{wt},\mathrm{Ras}}|}
{\tcontact}$, where $\xinitial^{\mathrm{wt},\mathrm{Ras}}$ denotes the position of 
each cell front at $t = 0$ (Fig.~\ref{fig02}).
The normal and transformed monolayers migrate with similar front velocities: 
$\V^{\mathrm{wt}}=16 \pm 5 \, \mu \mathrm{m \, h}^{-1}$ and 
$\V^{\mathrm{Ras}}=18 \pm 6 \, \mu \mathrm{m \, h}^{-1}$ (SD, $n=13$). 
We also measure the gap closure velocity, defined as $\V^{\mathrm{gap}}= \frac{\Delta x}{\tfusion} = 18 \pm 5 \, \mu\mathrm{m \, h}^{-1}$ 
(Fig.~\ref{fig03},a), consistent with the other definitions of the front velocity. 
We have checked that variations of the initial cell densities, and
of the initial front velocities, of the two monolayers do not impact 
the behavior of the interface after the meeting.

\noindent  The velocity fields were computed using PIV on the phase contrast images for 
$x<\xcontact$ and $t<\tcontact$. We note that the orientation of the velocity streamlines
for the wild type cells is more uniform (Fig.~S4).
The normal population migrates in a more directed 
manner than the transformed one. We checked that the mean velocities along the $y$-direction are close to zero 
for the two populations.

\begin{figure}[ht!]
	\centering
	\includegraphics[]{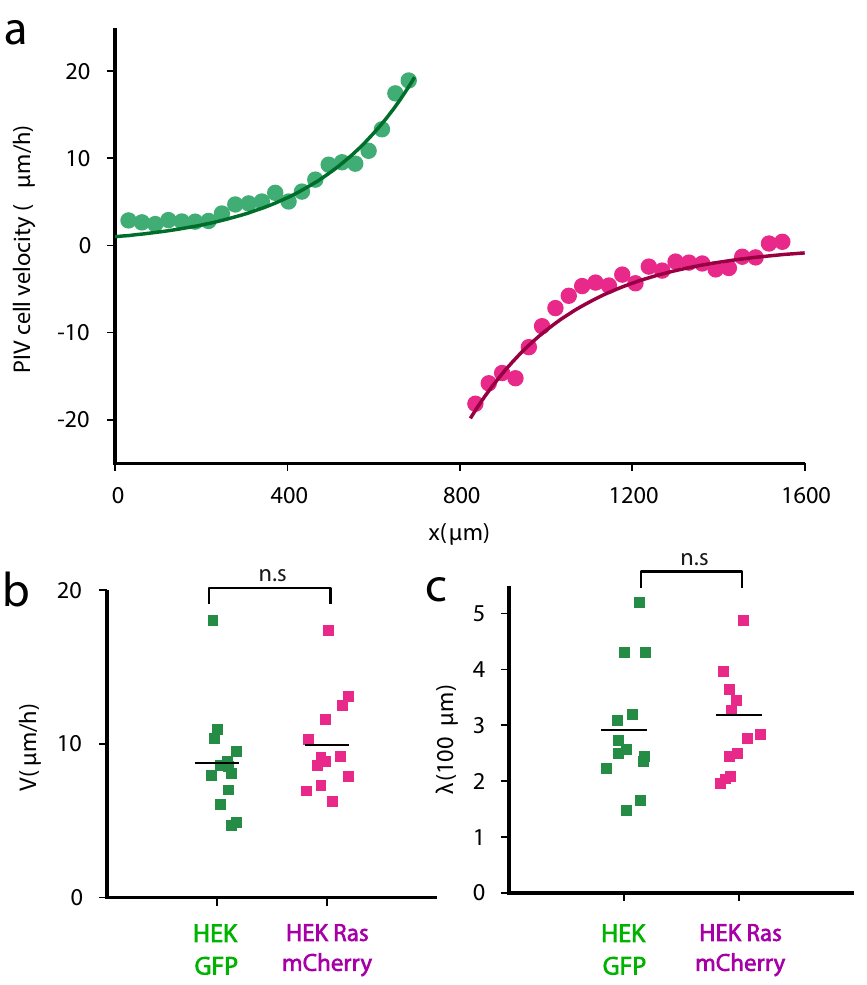}
	\caption{\textbf{Velocity profiles.} 
		a. Example of the AMA velocity profiles (dots) measured by PIV in the monolayers, $t = 6.75$ h. In this experiment $\tcontact = 8.5$ h.
		On the left hand side, the monolayer is composed of HEK-GFP cells (green dots) 
		and on the right hand side, the monolayer is composed of HEK-Ras-mCherry cells 
		(magenta dots). The green (resp. magenta) solid curve represents the best fit of 
		the function $V^{\mathrm{wt}}\, \exp((x-L)/\lambda^{\mathrm{wt}})$ 
		(resp. $-V^{\mathrm{Ras}}\, \exp(-(x-L)/\lambda^{\mathrm{Ras}})$), 
		where $L$ is the position of the front and $(V,\lambda)$ are the 
		fitting parameters. b. Set of fitted front velocities $V$. c. Set of fitted exponential decay lengths $\lambda$. Horizontal black lines correspond to mean values.
}
	\label{fig04}
\end{figure}

\begin{figure}[ht!]
	\centering
	\includegraphics[]{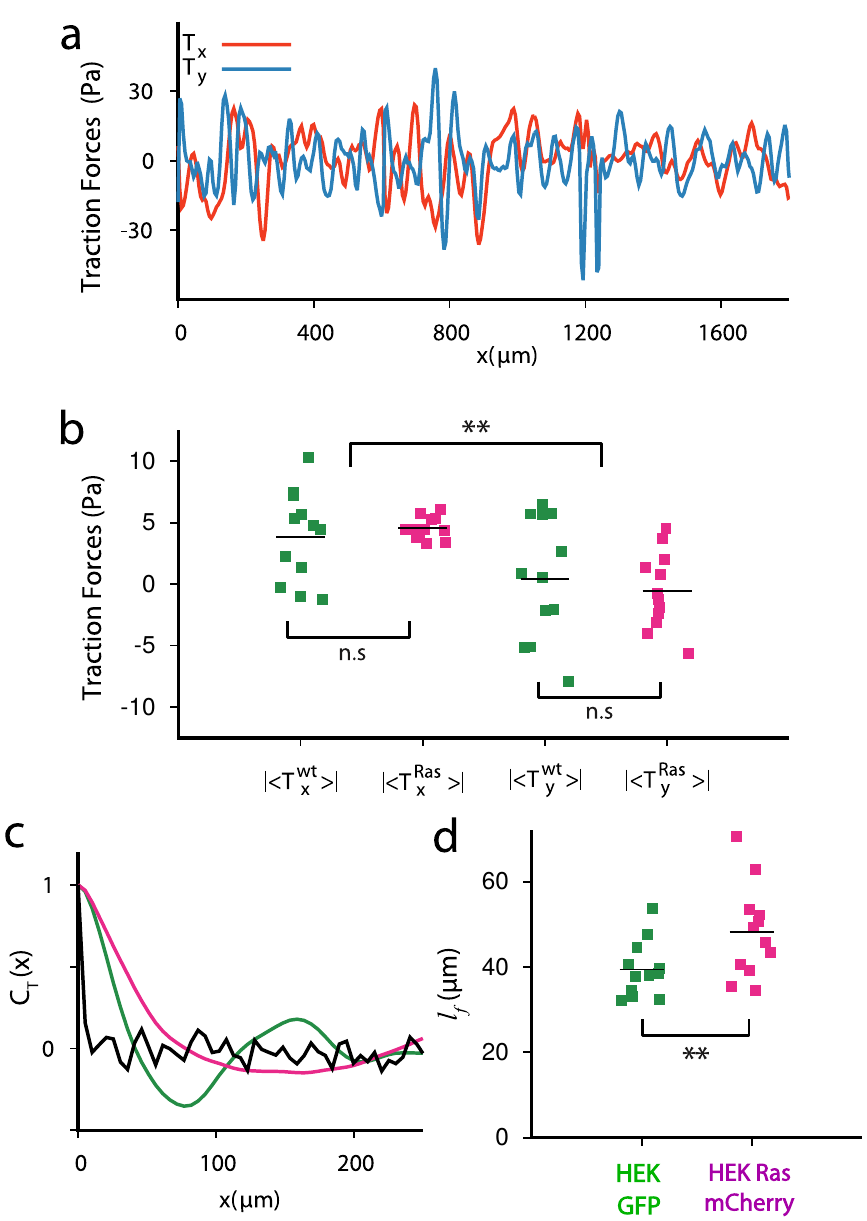}  
	\caption{\textbf{Characteristic parameters of traction force fields.}
		a. Traction force profiles $1$ h after barrier removal.
		b. Mean traction force components 
		$\langle T_x \rangle$ and $\langle T_y \rangle$
		for HEK-GFP (green squares) and HEK-Ras-mCherry 
		(magenta squares) cell monolayers. c. Autocorrelation functions of traction forces $T_x$ for HEK (green) and HEK Ras (magenta).
		The black curve is the autocorrelation function of white noise for control. 
		d. Correlation lengths of the $x$-component of traction forces $T_x$.
		All traction force data in the figure are averaged along the $y$ 
		direction, perpendicular to the direction of migration. Horizontal black lines correspond to mean values.
	}
	\label{fig05}
\end{figure}

\subsection{Data analysis}
For times before the first contact $t \le \tcontact$, we analyze
the velocity fields measured by PIV and the statistical properties of the 
traction force fields in the light of the theoretical framework given by 
Eqs.~(\ref{eq:def:model:1:const}-\ref{eq:def:model:1:polarity}). 
As shown in Fig.~\ref{fig04}a, the velocity profiles 
decay over a lengthscale $\lambda$ of several hundred micrometers
from a maximal value observed at the front.
Further, the velocity profiles are in good agreement with a \emph{single}
exponential function $\sim \pm V\, \exp( \pm(x-L(t))/\lambda)$
(Fig.~\ref{fig04},a). For times
  $\tcontact-2 \, \mathrm{h} \le t  \le \tcontact$, we
  checked that the fitting parameters $(V,\lambda)$ remained
  constant within error bars (Fig.~S5).

\noindent Cell traction force fields ($T_x,T_y$) display a 
rather noisy distribution in space without clear regular patterns 
(Fig.~\ref{fig05},a).
  We next focused on the statistical properties of collective cell traction
  forces including the free boundary, computing averages as explained above,
    but over windows which, for each monolayer, include their leading edges
  (see Characterization of cell lines). 
Upon averaging over the $y$-axis, we find that 
$|\langle T_x\rangle^{\mathrm{Ras,wt}}| > |\langle T_y\rangle^{\mathrm{Ras,wt}}|$
(Fig.~\ref{fig05},b), suggesting that the mean traction forces approximately 
parallels the direction of migration $x$. Unlike for assemblies of randomly 
oriented force dipoles, the mean traction forces $\langle T_x\rangle$ are 
non-vanishing for both cell types (Fig.~\ref{fig05},b), indicating that 
these cells coordinate forces over 
distances that are large compared to the typical cell size, 
as found for epithelial cells \cite{Trepat2009}.
We denote $l_f$ the decay length of the 
autocorrelation function of the $x$ component of the traction force
(Fig.~\ref{fig05},c).
We find that transformed cells coordinate force over longer 
distances $l_f^{\mathrm{Ras}} = 48\pm11 \, \mu\mathrm{m}$ than normal cells 
  $l_f^{\mathrm{wt}} =  39\pm7 \, \mu\mathrm{m}$ (SD, $n=12$) (Fig.~\ref{fig05},d).

\noindent Identifying $l_f$ with $L_c$, the comparison of the typical length 
scale of velocity variations ($\sim 300$~$\mu$m) with the correlation length 
of traction forces ($\sim 50$~$\mu$m), suggests that monolayer spreading  
occurs in the theoretical limit $L_c\ll L_{\eta}$, that we assume from now on. 
In this limit, Eq.~\eqref{eq:sol:v:1} reduces to a single exponential 
function $\sim  V\, \exp((x-L(t))/\lambda)$, 
where  $V$ is identified with the front velocity $\V$ and $\lambda$ 
with the hydrodynamic length $L_\eta$, while $\Sig \approx  T_0 L_c $ 
according to Eq.~\eqref{eq:def:sfront:1}. 
We estimate the parameters $V$ and $\lambda$ from the first two moments 
of the velocity profiles (see Materials and Methods), and deduce 
values of the hydrodynamic screening lengths  
$L_\eta^{\mathrm{Ras}}=320\pm110 \, \mu\mathrm{m}$ 
and $L_\eta^{\mathrm{wt}}=290\pm110 \, \mu\mathrm{m}$ (SD, $n=13$) 
(Fig.~\ref{fig04},c) that are large compared to the traction force correlation 
lengths (Fig.~\ref{fig05},d). Finally, we find that the front 
velocity of transformed monolayers is similar to that of normal ones: 
$\V^{\mathrm{Ras}}=9.9\pm3 \, \mu\mathrm{m \, h}^{-1}$
and $\V^{\mathrm{wt}}=8.7\pm3 \, \mu\mathrm{m \, h}^{-1}$  (SD, $n=13$) 
(Fig.~\ref{fig04},b). Note that the velocity amplitudes obtained by PIV 
are reduced by a factor of $~2$ compared to the estimates obtained from 
front displacements, which may be due to uncertainties of PIV 
techniques applied close to a free  boundary with a time-dependent fluctuating shape \cite{Deforet2012}.

\subsection{After contact, the normal monolayer moves backwards}

After the gap closes, the migration does not come to a halt and a competition 
for space arises between the two populations. The Ras monolayer continues 
to advance, while the wt population moves backwards. 
Although the details of the movements of the interface may vary from experiment to experiment, we always observe
the same direction of interface motion. Close to the 
interface, some cells from each population locally penetrate the opposite 
one, but the two populations essentially remain separated after fusion, 
thus forming a visible boundary between the two populations 
(Fig.~\ref{fig02}). To quantify the backward 
migration of the wt population, we measured the displacement 
$\dfront$ of the interface  separating the two populations during 
$\Delta t = 30$ h after contact. We found
$\dfront  =  83 \pm 50 \,\mu \mathrm{m}$ (SD, $n=13$)
with a variation range from a few micrometers
(almost static interface) to values larger than $100 \, \mu\mathrm{m}$. 
The speed of the interface $\VF = \dfront/\Delta t$ 
was deduced from this displacement,
$\VF=  2.7 \pm 1.7 \, \mu\mathrm{m \, h}^{-1}$  (Fig.~\ref{fig03},b).

\subsection{Estimation of the relative material parameters}

\begin{figure}[ht!]
	\centering
	\includegraphics[]{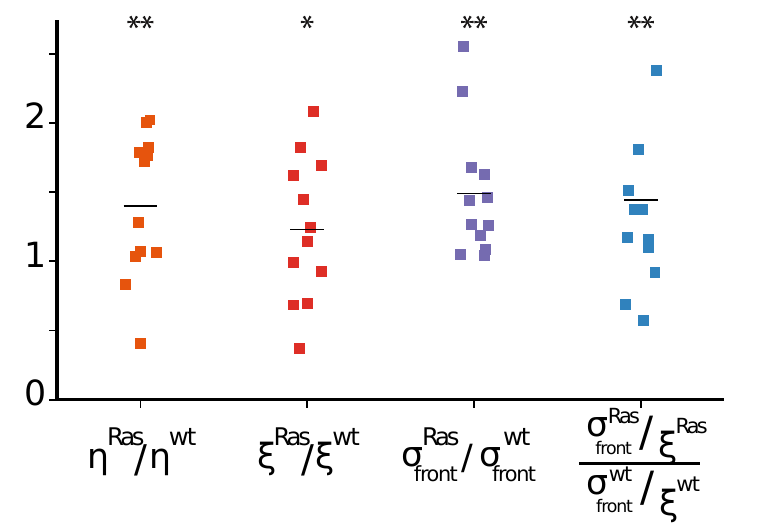}
	\caption{
\textbf{Relative material parameters for the two 
competing monolayers} From left to right, we show the relative 
shear viscosity $1.4\pm 0.5$, the relative friction coefficient $1.2\pm 0.5$, 
the relative front stress $1.5\pm 0.5$ (Eq.~\eqref{eq:def:sfront:1})  
and the relative ratio between front stress and friction coefficient 
$1.4\pm 0.7$ (SD, $n=12$). Horizontal black lines correspond to mean values. The stars refer to the p-value of a t-test comparing the ratios with $1$.
	}
	\label{fig06}
\end{figure}

The measurement of $\VF$ allows us to estimate the relative values of 
the material parameters of transformed and normal cell monolayers
in the light of the theoretical framework given by 
Eqs.~(\ref{eq:def:model:2:const}-\ref{eq:def:model:2:polarity})
(Fig.~\ref{fig06}). We use the values of $\V$ obtained from monolayer 
displacements, instead of the PIV values. We checked that
the fitting parameter $\lambda$ remained
constant for times $\tcontact \le t \le \tcontact+2\, \mathrm{h}$ 
(in agreement with the model hypothesis) whereas $V$ decreased
as expected towards $\VF$ (Fig.~S5).

First, we use Eq.~\eqref{eq:def:vfront:2} to obtain the ratio between 
the viscosities: $\eta^{\mathrm{Ras}}/\eta^{\mathrm{wt}}=1.4\pm0.5$ (SD, $n=12$).
Given the hydrodynamic lengths $L_{\eta}^{\mathrm{wt}}$,
$L_{\eta}^{\mathrm{Ras}}$, Eq.~(\ref{eq:def:Ln}), we next deduce the friction coefficients $\xi^{\mathrm{Ras}}/\xi^{\mathrm{wt}}=1.2\pm0.5$ (SD, $n=12$).
By combining these results with Eq.~\eqref{eq:def:sfront:1}, we estimate 
the ratio of the collective stresses at the front for both monolayers: 
$\Sig^{\mathrm{Ras}}/\Sig^{\mathrm{wt}}=1.5\pm 0.5$ (SD, $n=12$). In this sense, Ras-transformed cells are collectively stronger 
than normal cells. We conclude that the competition between the two cell 
populations can be framed as the dynamics of a moving interface between 
two compressible fluids with different front stresses.

\section{Discussion}

We interpret velocity measurements in antagonistic migration assays (AMAs) 
between wild type and Ras-transformed HEK cell sheets  in the
framework of a model in which the monolayers are considered as compressible and active
materials with different material parameters. Our analysis shows 
that collectively, transformed cells are characterized by a larger 
hydrodynamic length $L_{\eta}$, viscosity $\eta$, and cell-substrate 
friction coefficient $\xi$ than normal cells. 
Our model predicts that the direction of front migration is 
determined 
by the collective forces that build up at the fronts (\Sig), rather than 
by the traction force amplitude ($T_0$). 

\noindent Indeed, the average traction force amplitudes of both isolated cells and homogeneous monolayers are larger for normal than
for transformed cells. Although large variations of front and interface
positions make it hard to directly estimate $\Sig$ from the traction force 
data, we find that the ratio of the average component of traction
forces parallel to the direction of migration is consistent with $1$, 
$|\langle T_x\rangle^{\mathrm{wt}}|/|\langle T_x\rangle^{\mathrm{Ras}}|=0.85\pm0.82$
(SD, $n=12$), whereas the traction force correlation length is larger in the
transformed  monolayer compared to the wild type one, 
$l_f^{\mathrm{Ras}}/l_f^{\mathrm{wt}} =  1.25 \pm 0.31 $ (SD, $n=12$).
In this sense, Ras-transformed cells may collectively 
exert stronger front stresses than normal cells, 
$\Sig^{\mathrm{Ras}} > \Sig^{\mathrm{wt}}$. We emphasize that, at
the single cell level, Ras cells exhibit lower traction force amplitudes 
(Fig.~S2), whereas at the multicellular level both cell types 
exhibit forces of similar amplitude in bulk (Fig.~S3).
Determining how the collective mechanical properties of a cell assembly
emerge from individual cell properties and cell-cell interactions remains
an essential, but largely unsolved question.

\noindent Further, we verified that the ratio 
$(\Sig/\xi)^{\mathrm{Ras}}/(\Sig/\xi)^{\mathrm{wt}}=1.4\pm0.7$ (SD, $n=12$) 
(Fig.~\ref{fig06}) is larger than $1$, as already found in the
analysis of the kinematics of disk-shaped wound-healing assays with the
same cell lines \cite{Cochet-Escartin2014}. 
The quantitative discrepancy between the two (model-dependent) estimates of
$(\Sig/\xi)^{\mathrm{Ras}}/(\Sig/\xi)^{\mathrm{wt}}$
may arise due to different model hypotheses, as the monolayer
flow was assumed to be inviscid and incompressible in our previous work 
\cite{Cochet-Escartin2014}.

\noindent Interestingly, AMAs between normal and Ras$^{V12}$ MDCK cells \cite{Porazinski2016} 
show the opposite result (Ras MDCK cells being displaced backwards, while normal MDCK cells continue to migrate forward). 
In this work, the authors concluded that MDCK-Ras cells repulsion by normal MDCK cells is a process
that is dependent on E-cadherin-based cell-cell adhesion. In the present study, however, immunostaining for E-cadherin
revealed the absence of this protein at the cell-cell junctions for both normal and Ras-transformed HEK cells (Fig.~S6). Since E-cadherin is required for EphA2 receptor localization
at cell-cell contacts \cite{Zantek1999,Orsulic2000}, Eph receptor signaling cannot be directly
involved in our system. On the basis of the present analysis,
we conjecture that collective stresses are stronger in MDCK wt cell sheets
compared to MDCK Ras cell sheets. Irrespective of the cell line, the 
connection between molecular constituents and their respective 
expression levels in normal and transformed cells on the one hand,
and the respective hydrodynamic parameter values on the other hand,
remains unknown and deserves further study.

\noindent In our theoretical framework we have omitted several effects that might be relevant for AMAs, like specific molecular interactions between the two cell populations or changes of cell polarity after contact. Indeed, cell behavior is known to be influenced by the local micro-environment, and thus leading cells may actively change their orientation and repolarize upon fusion with the competing tissue. If confirmed by observation, this effect could be taken into account by changing accordingly the boundary conditions for the polarity fields Eqs.~\eqref{eq:def:BC:2:sig}, which would lead to \V ~being weighted differently in Eq.~\eqref{eq:def:vfront:2}, and to different values of the model-dependent relative parameters.
Over longer time scales, tissue material parameters may become time-dependent
\cite{Blanch-Mercader2017,Garcia2015}, and differences in cell proliferation rates
may become relevant \cite{Ranft2014}. Since we focused here on the vicinity 
of the contact time between the two populations, we defer to future work the
incorporation of these additional ingredients into our theoretical framework.

\noindent Our analysis illustrates that AMAs can be used to estimate relative hydrodynamic parameters of spreading monolayers from their kinematics only. We believe that this setting is a useful testing ground to explore the mechanisms governing competition between cellular assemblies.

\section*{Conflicts of interest}
There are no conflicts to declare.

\section*{Acknowledgements}
HEK cells are a gift from M.C.~Parrini, Institut Curie, France. We thank V.~Hakim and the members of the team ``Biology-inspired physics at mesoscales'' for fruitful discussions. The authors belong to the CNRS research consortium (GdR) 'CellTiss'. This work was funded by La Ligue (\'Equipe labellis\'ee), the Labex CelTisPhyBio, the GEFLUC Ile-de-France and the C'Nano Ile-de-France (projet COMPCELL). SM was supported by a doctoral fellowship from the IPGG, KS by a FPGG grant, and TM was funded by Ecole de l'INSERM.

\section{Appendix A}

In this Appendix, we solve  the evolution equations 
(\ref{eq:def:model:1:const}-\ref{eq:def:model:1:polarity})
with the boundary conditions:
\begin{eqnarray}
\label{eq:def:BC:2}
v(x = L) &=& 0 \,,\\
p(x = L) &=& +1 \,,
\end{eqnarray}
valid when a single cell sheet located in the fixed domain $x < L$ is
clamped at position $x=L$.
The polarity profile is unchanged, see 
Eq.~\eqref{eq:sol:p}. However the velocity and stress profiles now read:
\begin{eqnarray}
\label{eq:sol:v:2}
v(x) &=& \frac{ \V L_c}{L_c - L_{\eta}} 
\left(\exp \left( \frac{x - L}{L_c}\right)
-\exp \left( \frac{x - L}{L_{\eta}}\right)
\right)\,, \\
\sigma(x) &=&\frac{ \eta \V L_c}{L_c - L_{\eta}} 
\left(\frac{1}{L_c} \, \exp \left( \frac{x - L}{L_c}\right)
-\frac{1}{L_\eta} \, \exp \left( \frac{x - L}{L_{\eta}}\right)
\right)\,.
\end{eqnarray}
The maximal stress is applied by the monolayer at the front, 
with $\Sig = -\sigma(x = L, t)$ given by \eqref{eq:def:sfront:1}.

% \bibliographystyle{unsrt}
% \bibliography{competition} 

\newpage
\setcounter{figure}{0}
\setcounter{table}{0}

\pagestyle{empty}

\section*{Supplementary Information}

\textbf{Supplementary Movie 1:} A typical AMA between HEK-GFP wild type cells (green) and HEK Ras cells (magenta) showing the backward migration of the GFP population after meeting.The time reference $t=0$ is set when the physical barrier is removed. Scale bar : $150 \,\mu \mathrm{m}$.

\bigskip
\noindent
\textbf{Supplementary Figures:}
\medskip
\begin{figure}[!h]
\centering
  \includegraphics[height=6cm]{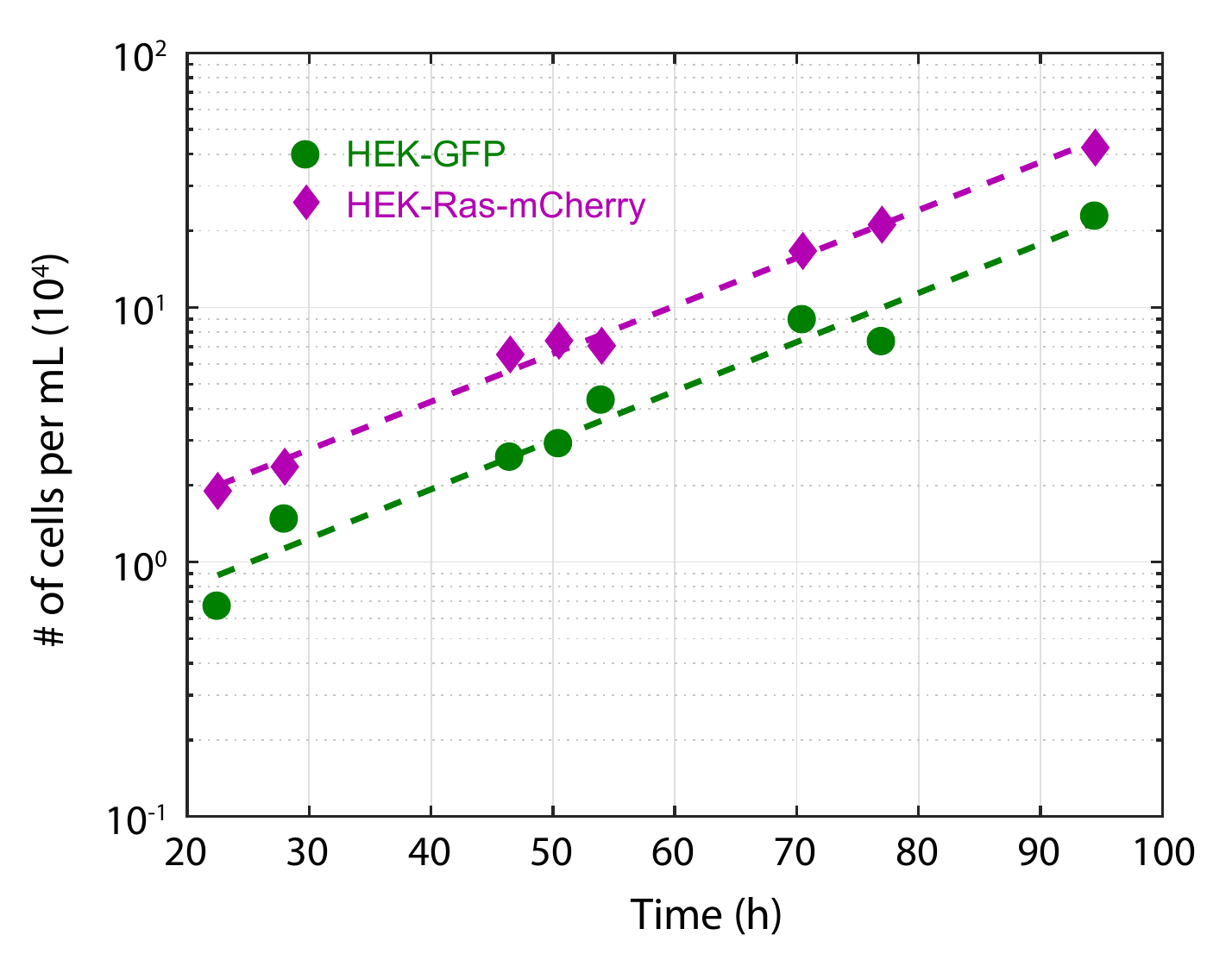}
  \captionsetup{labelformat=suppfig}
	\caption{\textbf{Estimation of the population doubling time.} 
Semi-logarithmic graph of the cell number \emph{vs.} time, for the
HEK-GFP (green circles) and HEK-Ras-mCherry (magenta diamonds) cell populations. Slopes of the dashed lines give an estimation of the population doubling time $\tau_d = 16$ h.
}
  \label{figS1}
\end{figure}

\begin{figure}[!h]
\centering 
\includegraphics[height=7cm]{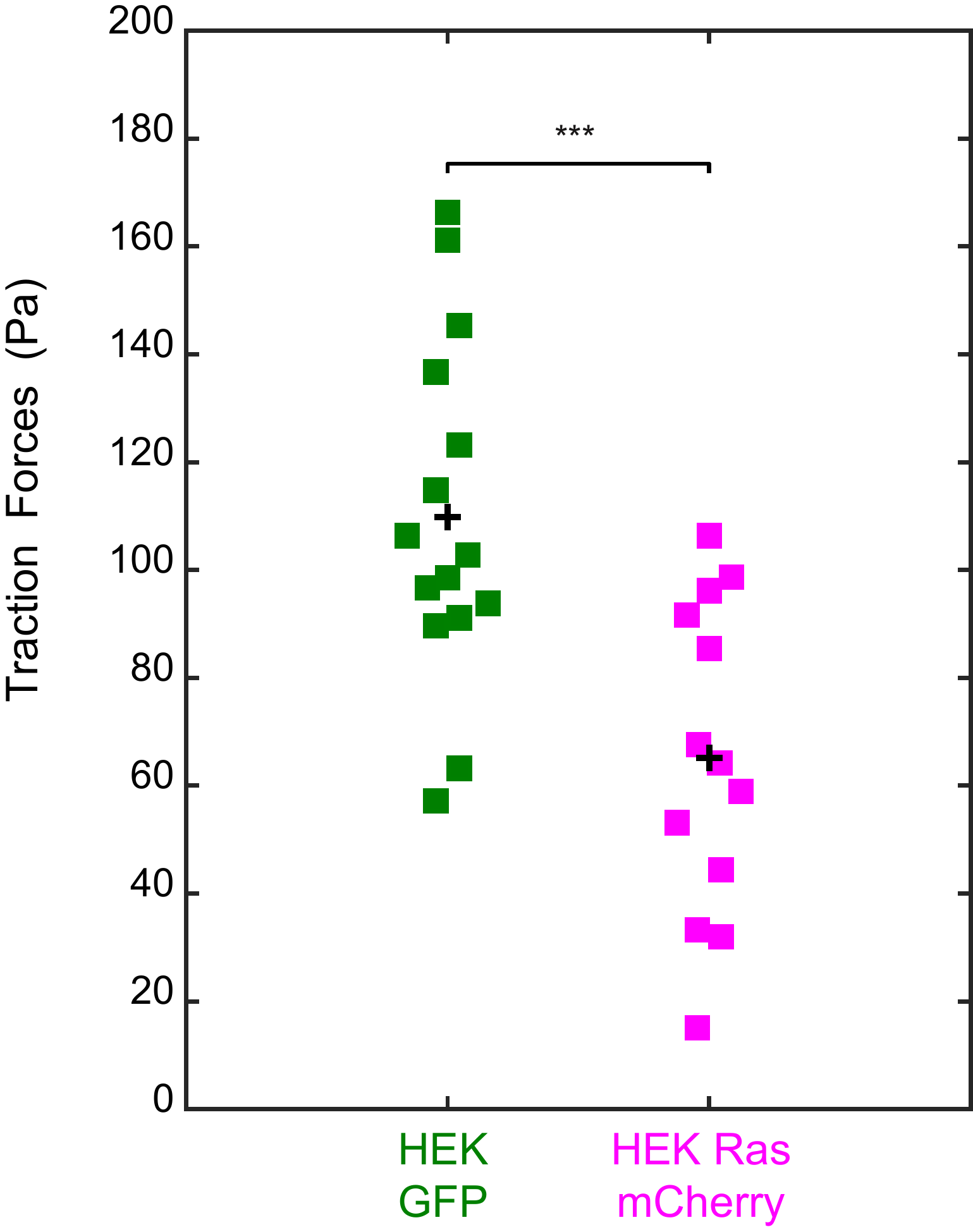}
\captionsetup{labelformat=suppfig}
  \caption{\textbf{Single cell traction forces.}
    Average traction force amplitudes (Pa)
of isolated adherent HEK cells. HEK wt cells (green) exert higher traction 
forces on the substrate than HEK Ras cells (magenta).}
  \label{figS2}
\end{figure}

\begin{figure}[!h]
	\centering 
	\includegraphics[width=8.6cm]{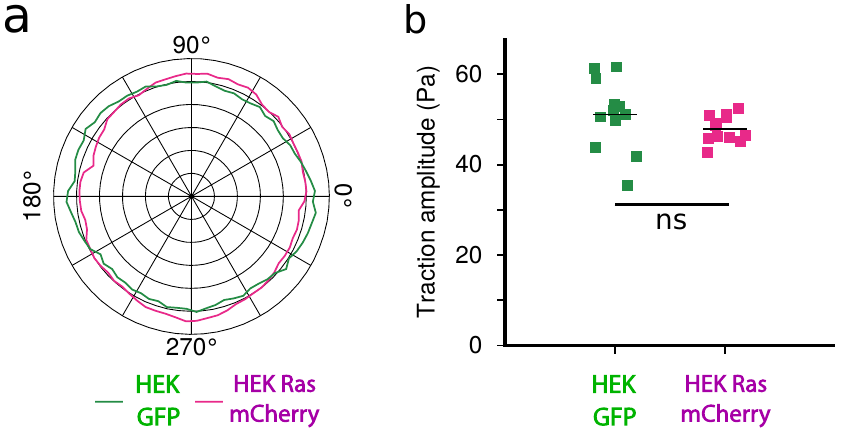}
	\captionsetup{labelformat=suppfig}
	\caption{
\textbf{Bulk collective cell traction forces.}
		Traction force measurements of confluent monolayers between $1~\mathrm{h}$ to $3.5~\mathrm{h}$ after barrier removal. Only  HEK wt cells (green) and HEK Ras cells (magenta) that are at least $700~\mu$m away from the corresponding leading edges were considered. (a) Distribution of force orientation. (b) Average amplitude of traction forces. Values are substantially larger than in Fig.~\ref{fig05},b where average traction force \emph{components} are plotted.
}
	\label{figS3}
\end{figure}

\begin{figure}[!t]
\centering 
\includegraphics[height=5cm]{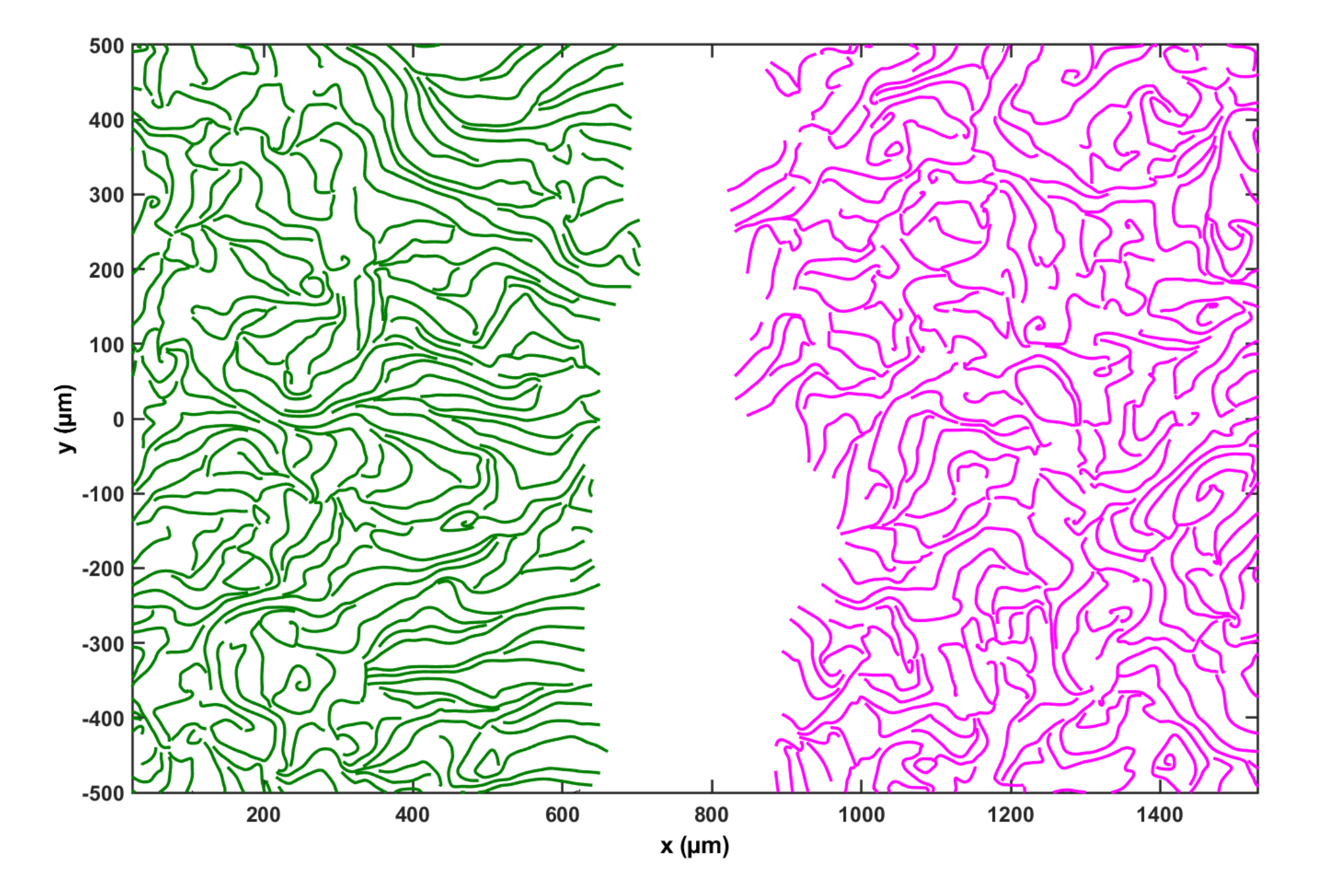}
\captionsetup{labelformat=suppfig}
  \caption{\textbf{Flow field of the AMAs before contact.} Streamlines of the velocity fields for HEK wt cells (green) and HEK Ras cells (magenta) at $t=6.6$~h of an AMA. Note that the streamlines for wild-type cells are more ordered at the front than for HEK Ras cells.}
  \label{figS4}
\end{figure}

\begin{figure}[!t]
	\centering 
	\includegraphics[width=8.6cm]{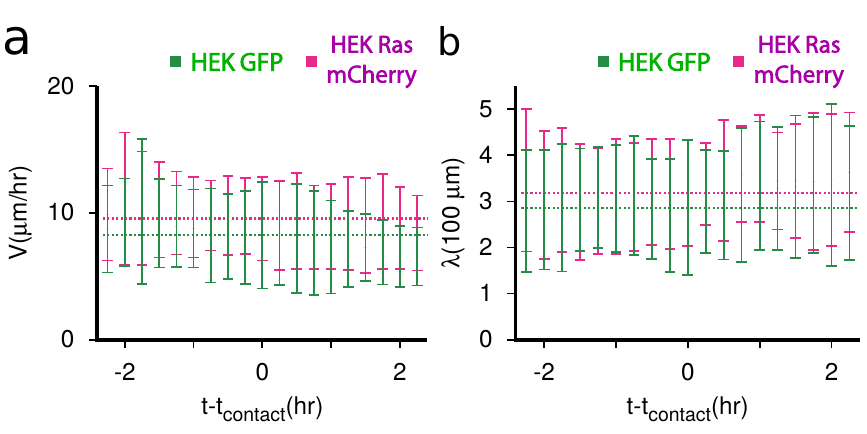}
	\captionsetup{labelformat=suppfig}
	\caption{
\textbf{Time dependence of the fitting parameters.}
		The AMA velocity profiles at $\tcontact-2~\mathrm{h}<t<\tcontact+2~\mathrm{h}$ are fitted with the functions $V^{\mathrm{wt}}\, \exp((x-L)/\lambda^{\mathrm{wt}})$ and $-V^{\mathrm{Ras}}\, \exp(-(x-L)/\lambda^{\mathrm{Ras}})$, where $L$ is the position of the front and $(V,\lambda)$ are the fitting parameters (see Fig.~\ref{fig04}). a. Front velocity $V$ as a function of time. b. Exponential decay length $\lambda$ as a function of time. Green corresponds to HEK wt cells and magenta to HEK Ras cells. Dashed lines correspond to the mean values at $\tcontact$. Error bars denote standard deviations.
}
	\label{figS5}
\end{figure}

\begin{figure}[!t]
\centering 
\includegraphics[height=9cm]{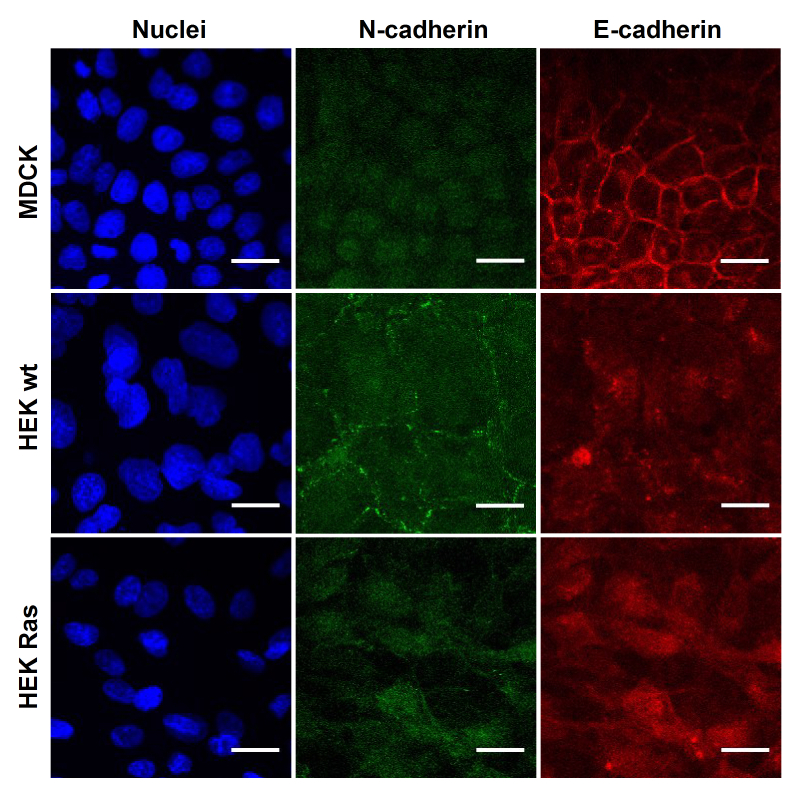}
\captionsetup{labelformat=suppfig}
  \caption{\textbf{Immunostaining for cell-cell junctions.} MDCK, HEK wt and HEK Ras monolayers were fixed and stained with N-cadherin antibody (green) and E- cadherin antibody (red) and Dapi for nuclei (blue). E-cadherin
is rather weak and localised through entire cells for both HEK cells. N-cadherin is also cytoplasmic for HEK ras cells while being localized at the cell-cell junctions for HEK normal cells. Scale bars = 20~$\mu$m.}
  \label{figS6}
\end{figure}

\end{document}